\documentclass[fleqn,twoside]{article}
\usepackage{espcrc2}
\input epsf

\usepackage{graphicx}
\usepackage[figuresright]{rotating}

\newcommand{\AmS}{{\protect\the\textfont2
  A\kern-.1667em\lower.5ex\hbox{M}\kern-.125emS}}

\title{Status and preliminary results of the ANAIS experiment at Canfranc}

\author{S. Cebri\'{a}n\address{Laboratory of Nuclear and High Energy Physics, University of
Zaragoza, 50009 Zaragoza, Spain}\thanks{Attending speaker:
scebrian@posta.unizar.es}, J. Amar\'{e}\addressmark, J. M.
Carmona\addressmark, E. Garc\'{\i}a\addressmark[Zaragoza], I. G.
Irastorza\addressmark\thanks{Present address: CERN, EP Division,
CH-1211 Geneva 23, Switzerland}, G. Luz\'{o}n\addressmark, A.
Morales\addressmark, J. Morales\addressmark, A. Ortiz de
Sol\'{o}rzano\addressmark, J. Puimed\'{o}n\addressmark, M.L.
Sarsa\addressmark, J. A. Villar\addressmark }

\begin{document}

\begin{abstract}
ANAIS (Annual Modulation with NaI's) is an experiment planned to
investigate seasonal modulation effects in the signal of galactic
WIMPs using up to 107 kg of NaI(Tl) in the Canfranc Underground
Laboratory (Spain). A prototype using one single crystal (10.7 kg)
is being developed before the installation of the complete
experiment; the first results presented here show an average
background level of 1.2 counts/(keV~kg~day) from threshold
($E_{thr} \sim 4$ keV) up to 10 keV.
 \vspace{1pc}
\end{abstract}

\maketitle

\section{INTRODUCTION}

There is a substancial evidence to conclude that most of the
matter in the Universe must be dark and that it consists mainly of
cold non-baryonic particles. Weak Interacting Massive Particles
(WIMPs) are favourite candidates to such non-baryonic components.
A convicing proof of the detection WIMPs, which are supposedly
filling the galactic halo, would be to find unique signatures in
the data, like seasonal asymmetries. ANAIS (Annual Modulation with
NaI's) is a large mass experiment intended to investigate the
annual modulation effect which would be produced in the signal of
galactic WIMPs due to the va\-ria\-tions in the relative velocity
between the Earth and the halo \cite{Freese88}. It will be
installed in the Canfranc Underground Laboratory, located in an
old railway tunnel in the Spanish Pyrenees with an overburden of
2450 m.w.e., using up to 10 NaI(Tl) hexagonal crystals (10.7 kg
each) as an improved scaled-up version of a previuos experiment
\cite{Sarsa97}. Before setting-up the whole experiment, a
prototype is being developed in Canfranc in an attempt to obtain
the best energy threshold and lowest radioactive background in the
low energy region (2 to 50 keV), as well as to check the stability
of the environmental conditions which influence on the detector
response.

\section{THE ANAIS PROTOTYPE}

One single detector has been used in the ANAIS prototype; it
consists of a NaI(Tl) crystal encapsulated inside 0.5-mm-thick
stainless steel and coupled to a PMT through a quartz window. Some
components of the photomultiplier have been removed because of
their radioimpurities. The scintillator has been placed in a
shielding consisting of 10 cm of archaeological lead (of less than
9~mBq/kg of $^{210}$Pb) followed by 20 cm of low activity lead, a
sealed box in PVC (maintained at overpressure to prevent the
intrusion of radon), 2-mm-thick cadmium sheets, and finally, 40 cm
of polyethylene and tanks of borated water. An active veto made of
plastic scintillators is covering the set-up.

The data acquisition system, based on standard NIM and CAMAC
electronics, has two diffe\-rent parts following the two output
signals implemented from the PMT; the fast signal is recorded
using a digital oscilloscope while the slow signal is routed
through a linear amplifier and analog-to-digital converters
controlled by a PC through parallel interfaces, to register the
energy of events up to $\sim 1.7$~MeV.

Parameters of the pulses used to reject the noise produced by PMT
in the various previous NaI experiments are the mean amplitude
\cite{Gerbier99}, a ratio of area portions \cite{Bernabei99}, etc.
In the present work the filtering of noise uses the squared
deviation of the digitalized pulse from the well-known
theo\-re\-ti\-cal shape of a scintillation event of the same area.
To reject the noise, a safe cut at $3\sigma$ from the center of
the gaussian distribution of this parameter for calibration events
from 4 to 10~keV has been used.

By comparing the data recorded from December 2000 to August 2001
with the Monte Carlo simulations, it was possible to identify the
main sources of background in the region of interest. The
$^{210}$Pb 46.5~keV line as well as a peak due to the escape of
X-rays of I at $\sim$ 16~keV seen in the spectrum, may be caused
by the presence of radioimpurities in the stainless steel can
and/or in the PMT. The area of the 1460.8~keV peak is compatible
with an activity of 15~mBq/kg from $^{40}$K in the NaI crystal;
these impurities produce an almost flat background in the low
energy region due to their beta emission.
A comparison between the spectra recorded with and without the
neutron shielding does not show noticeable differences.

A pulse shape analysis has been carried out with the purpose of
investigate the possible appea\-rance of the so-called "anomalous"
or "bump" events found in other NaI experiments
\cite{Liubarsky00,Gerbier00}.
No evidence of such anomaly has been found in the distributions
for background events, neither following the method of the UKDMC
(fitting integrated pulses to calculate the decay time cons\-tant)
nor using other parameters (like the first momemtum of the pulse).

Monitoring and stabilisation control of the environmental
conditions (radon levels, $N_{2}$ flux, temperature in the
laboratory and in the inner enceinte, photomultiplier working
voltage, \dots) is underway. Using the data of the prototype,
collected along almost 6000 hours, the stability of some
parameters has been checked. The fluctuations of the ADC channels
for the different peaks used to perform the energy calibration
range from 1 to 1.5 \%. With respect to the counting rates, the
gaussian distributions of the deviations from the mean values have
a sigma of 1.27 for the rate integrated above 6~keV and 1.47 for
the rate above 100~keV.

\section{FIRST RESULTS}

\begin{figure}[htb]
\centerline{ \fboxrule=0cm
 \fboxsep=0cm
  \fbox{
\epsfxsize=8cm
  \epsffile{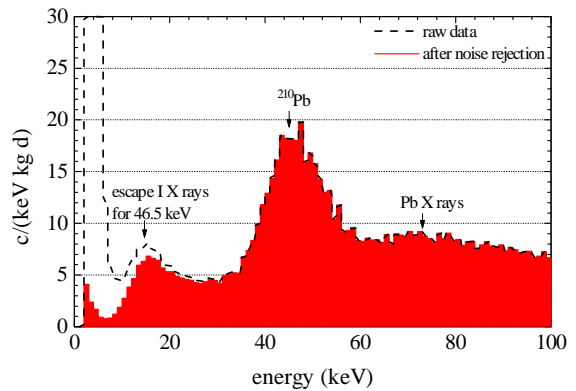}}}
\begin{center}
{\caption {Low-energy region of the observed spectrum in the
prototype of ANAIS before and after the noise
rejection.}\label{spectrum}}
\end{center}
\end{figure}

The results presented here correspond to an exposure of 1225.4
kg$\times$day. Fig. \ref{spectrum} shows the raw spectrum and the
spectrum after the noise rejection up to 100 keV.
The energy threshold is of $\sim 4$ keV and the background level
registered from the threshold up to 10~keV is about 1.2
counts/(keV~kg~day).

\begin{figure}[htb]
\centerline{ \fboxrule=0cm
 \fboxsep=0cm
  \fbox{
\epsfxsize=7cm
  \epsffile{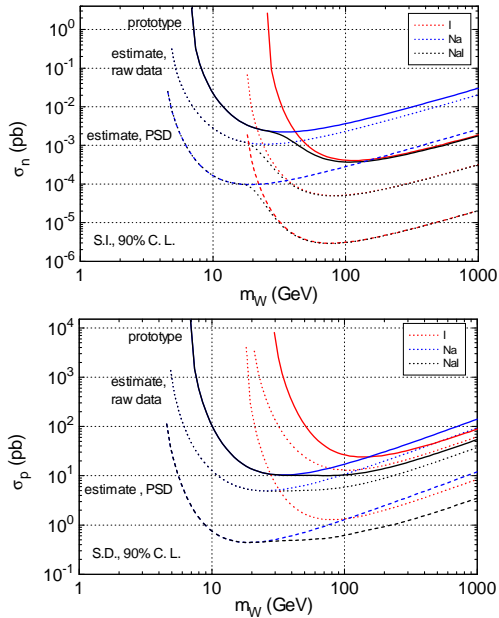}}}
\begin{center}
{\caption {Exclusion plots derived for SI (top) and SD (bottom)
interactions from the prototype (solid line) and expected for the
whole experiment with (dashed line) and without (dotted line) PSD
techniques.}\label{plot}}
\end{center}
\end{figure}

We have used this region to derive the corresponding limits for
the WIMP-nucleon cross sections. The galactic halo is supposed to
be isotropic, isothermal and non-rotating, assuming a density of
$\rho$=0.3 GeV/cm$^{3}$, a Maxwellian velocity distribution with
$v_{rms}$=270 km/s (with an upper cut corresponding to an escape
ve\-lo\-ci\-ty of 650 km/s) and a relative Earth-halo velocity of
$v_{r}$=230 km/s. The Helm parameterization \cite{Engel91} is used
for the coherent form factor, while the approximation from
\cite{Lewin96} is con\-si\-dered for the SD case. Spin factors
($\lambda_{p}J(J+1)$) 0.089 and 0.126 are assumed for Na and I
respectively. Fig. \ref{plot} shows, in addition to the limits
derived from the prototype results (solid lines), the estimates
considering a flat background of 1 count/(keV~kg~day) from 2 to 8
keV after an exposure of 107 kg$\times$y both for raw data (dotted
lines) and assuming PSD (dashed lines). The plots show the contour
lines for each nucleus, Na and from I, as well as the NaI case.
That is shown both for SI scalar interactions and SD WIMP-proton
interactions. It should be noted that for SI interactions and
using PSD techniques, ANAIS will be able to explore the region of
WIMPs singled out by the possible annual modulation effect
reported by the DAMA colla\-bo\-ra\-tion \cite{Bernabei00}.

\section{FUTURE PROSPECTS}

The next steps in the development of the prototype of ANAIS,
according to these first results, are the removal of the present
PMT and the steel can and to install, instead, two ultra-low
background PMT and a 1-cm-thick teflon enclosure filled with
special mineral oil, as in the NAIAD experiment \cite{Spooner00}.
A program of measurements to select high radiopurity materials is
in course in Canfranc, using an ultra-low background Ge detector.
The program includes the removal of components when neccesary to
reduce, as much as possible, the various sources of background, to
diminish the noise by using anticoincidence read-out (lowering so
also the energy threshold) and improving the collection of the
scintillation light.


\section*{Acknowledgements}
The Canfranc Astroparticle Underground La\-bo\-ra\-to\-ry is
operated by the University of Zaragoza under contract No.
AEN99-1033. This research was funded by the Spanish Commission for
Scien\-ce and Technology (CICYT) and the Government of Arag\'{o}n.

\end{document}